\documentclass[sigconf]{acmart} 
\AtBeginDocument{%
  }

\copyrightyear{2026}
\acmYear{2026}
\setcopyright{cc}
\setcctype{by-nc-nd}
\acmConference[CHI EA '26]{Extended Abstracts of the 2026 CHI Conference on Human Factors in Computing Systems}{April 13--17, 2026}{Barcelona, Spain}
\acmBooktitle{Extended Abstracts of the 2026 CHI Conference on Human Factors in Computing Systems (CHI EA '26), April 13--17, 2026, Barcelona, Spain}
\acmDOI{10.1145/3772363.3798475}
\acmISBN{979-8-4007-2281-3/2026/04}
\usepackage{subcaption}
\usepackage{diagbox}
\usepackage{hhline}
\usepackage{multirow}
\usepackage{colortbl}
\usepackage{graphicx}

\usepackage{algpseudocode}
\usepackage{tikz}
\usepackage{enumitem}
\usetikzlibrary{shapes.geometric, arrows, positioning, fit}

\begin{document}


\title{TinyGaze: Lightweight Gaze-Gesture Recognition on Commodity Mobile Devices}

% \author{anonymous}

\author{Yaxiong Lei}
\email{yl212@st-andrews.ac.uk}
\orcid{0000-0002-0697-7942}
\affiliation{%
  \institution{University of St Andrews}
  \city{St Andrews}
  \country{UK}
}
\affiliation{%
  \institution{University of Essex}
  \city{Colchester}
  \country{UK}
}

\author{Hyochan Cho}
\email{hc240@st-andrews.ac.uk}
\orcid{0009-0006-8048-6937}
\affiliation{%
  \institution{University of St Andrews}
  \city{St Andrews}
  \country{UK}
}

\author{Fergus Buchanan}
\email{fergusbuchanan9@gmail.com}
\orcid{0009-0008-5692-6183}
\affiliation{%
  \institution{University of St Andrews}
  \city{St Andrews}
  \country{UK}
}

\author{Shijing He}
\email{shijing.he@kcl.ac.uk}
\orcid{0000-0003-3697-0706}
\affiliation{%
 \institution{King's College London}
 \city{London}
 \country{UK}
}

\author{Xinya Gong}
\email{xg31@st-andrews.ac.uk}
\orcid{0009-0005-6414-9351}
\affiliation{%
  \institution{University of St Andrews}
  \city{St Andrews}
  \country{UK}
}

\author{Yuheng Wang}
\email{yw99@st-andrews.ac.uk}
\orcid{0000-0003-3335-8706}
\affiliation{%
  \institution{University of St Andrews}
  \city{St Andrews}
  \country{UK}
}

\author{Juan Ye}
\email{Juan.Ye@st-andrews.ac.uk}
\orcid{0000-0002-2838-6836}
\affiliation{%
 \institution{University of St Andrews}
  \city{St Andrews}
  \country{UK}
}

\renewcommand{\shortauthors}{Yaxiong Lei et al.}

\begin{abstract}
% Gaze gestures can provide hands free input on mobile devices, but practical use requires (i) gestures users can learn and recall and (ii) recognition models that are efficient enough for on device deployment. We present an end-to-end pipeline using commodity ARKit head/eye transforms and a scaffolded guidance-to-recall protocol grounded in learning theory. In a pilot study (4 participants; 240 trials), we benchmark a compact time-series model (TinyHAR) against deeper baselines (DeepConvLSTM, SA-HAR) on 5-way gesture recognition and 4-way user identification. TinyHAR achieves 0.960 Macro F1 for gesture recognition and 0.997 Macro F1 for user identification while using only 46k parameters. A modality analysis further shows that head pose dynamics can be highly informative for mobile gaze gestures, highlighting embodied head-eye coordination as a key design consideration.
Gaze gestures can provide hands free input on mobile devices, but practical use requires (i) gestures users can learn and recall and (ii) recognition models that are efficient enough for on-device deployment. We present an end-to-end pipeline using commodity ARKit head/eye transforms and a scaffolded guidance-to-recall protocol grounded in learning theory. In a pilot feasibility study (N=4 participants; 240 trials; controlled single-session setting), we benchmark a compact time-series model (TinyHAR) against deeper baselines (DeepConvLSTM, SA-HAR) on 5-way gesture recognition and 4-way user identification. TinyHAR achieves strong performance in this pilot benchmark (Macro F1 = 0.960 for gesture recognition; Macro F1 = 0.997 for user identification) while using only 46k parameters. A modality analysis further indicates that head pose dynamics are highly informative for mobile gaze gestures, highlighting embodied head--eye coordination as a key design consideration. Although the small sample size and controlled setting limit generalizability, these results indicate a potential direction for further investigation into on-device gaze gesture recognition.
% Given the small sample and controlled setting, these results should be interpreted as feasibility evidence rather than a generalizable deployment claim.
\end{abstract}





\begin{teaserfigure}
  \centering
  \includegraphics[width=1\textwidth]{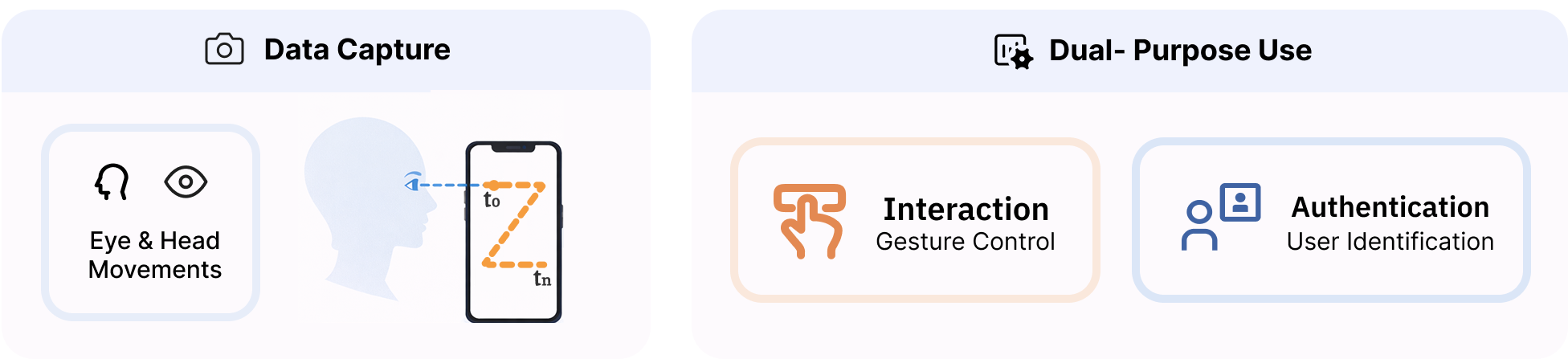}
  \caption{TinyGaze overview. TinyGaze logs ARKit head and eye pose time series and represents each gaze gesture as a spatio-temporal trajectory (from $t_0$ to $t_n$). The same sequential signal supports two tasks: (i) gesture control and (ii) continuous-style authentication via user identification.}
  % \Description{Overview diagram of TinyGaze showing ARKit head and eye pose streams forming a trajectory over time, used for gesture recognition and user identification.}
  \Description{An overview diagram of the TinyGaze system. The figure shows two stages: data capture and dual-purpose use. In the data capture stage, ARKit head and eye movement streams are recorded and represented as a time-ordered trajectory from t0 to tn on a smartphone screen. In the dual-purpose use stage, the same sequential signal is used for two downstream tasks: gesture-based interaction control and user identification for continuous-style authentication.}
  \label{fig:teaser}
\end{teaserfigure}

\maketitle

\section{Introduction}
Gaze is a powerful and natural input modality for Human-Computer Interaction~\cite{jacob1991use, zhai2003s, lei2023end} and a promising modality for privacy-preserving interaction~\cite{lei2023protecting, katsini2020role, he2025identity}. On mobile devices, gaze input provides a compelling hands-free alternative to touch. A key challenge in gaze interaction is the \textit{Midas Touch} problem~\cite{mohan2018dualgaze, lei2023dynamicread}, where the system cannot distinguish between a user's look to browse versus a look to select. Gaze gestures, represented as deliberate patterns of eye movement, can be an effective solution to this problem~\cite{drewes2007interacting, elmadjian2021gazebar, lei2023dynamicread}.

Beyond recognition accuracy, practical gaze gestures also depend on \emph{learnability}~\cite{aoki2008learning}: users must be able to reliably acquire and recall gesture patterns. Inspired by learning theory (scaffolding and faded guidance)~\cite{van2010scaffolding,roediger2011critical}, we structure gesture performance into stages that progressively remove visual support (guided tracing $\rightarrow$ outline cue $\rightarrow$ partial cue $\rightarrow$ free recall), encouraging robust motor/visuomotor memory rather than one-off imitation.

However, for gaze gestures to be practical on mobile devices, the recognition models have to be both accurate and computationally efficient. Much research has focused on complex deep learning models~\cite{lei2023end, shi2021gaze, khamis2018past}, which may not be suitable for real-time, on device processing. Accordingly, our goal is to pair a learnability-oriented protocol with lightweight recognition that is realistic for on device use. To address this, we investigate the following research question: \textbf{RQ:} Is it feasible to deploy lightweight models on mobile devices that match the accuracy of deeper baselines for gaze gestures while minimizing latency?

Our contributions include: (1) introduce an ARKit-based mobile sensing and preprocessing pipeline for gaze-gesture modelling, (2) propose a learning theory-inspired protocol that transitions from guided tracing to memory-based recall, and (3) provide a pilot benchmark comparing compact vs.\ heavier deep models on gesture recognition and user identification, including a modality analysis (head vs.\ eyes vs.\ fused).

\section{Related Work}
Classic gaze interfaces often rely on dwell-time, confirmations, or multimodal coupling (e.g., gaze + touch/head) to disambiguate attention from intent~\cite{jacob1991use, zhai2003s, mohan2018dualgaze, lei2023dynamicread, kong2021eyemu}. While effective, these mechanisms can add latency (dwell), workload (confirmations), or require extra modalities that are unavailable in hands-free contexts. Gaze gestures provide a complementary solution by encoding intent as a deliberate spatiotemporal pattern, enabling fast and expressive commands without external input channels~\cite{drewes2007interacting, elmadjian2021gazebar}.

Gaze gesture recognition has used template-based methods such as the \$1 recognizer~\cite{wobbrock2007gestures}, which are simple and interpretable but sensitive to noise and timing variation~\cite{lei2025quantifying}. Probabilistic sequence models (e.g., HMMs) capture temporal structure but rely on feature engineering and can struggle with high-dimensional raw signals~\cite{fujii2018gaze}. Classical ML (e.g., SVMs) has also been applied to gaze-derived classification~\cite{zhang2016gender}. More recently, deep learning (CNN/RNN hybrids and attention-based models) has achieved strong accuracy on gaze time series, especially with large datasets or on device IMU sensing~\cite{palmero2018recurrent, shi2021gaze, lei2023end, lei2025mac}. However, these models may be computationally expensive and can overfit when data are limited or sensor noise is high.

Compared to dedicated eye trackers, mobile gaze estimation via ARKit/TrueDepth is noisier and subject to drift, partial occlusion, and motion artifacts~\cite{parker2022eye}, making robustness and on device efficiency especially important for gesture input. This motivates the utilisation of compact deep learning architecture on time series data such as TinyHAR~\cite{zhou2022tinyhar}.

Beyond interaction, gaze and head motion can encode user-specific patterns for identification and continuous authentication~\cite{he2025identity}. Prior work on mobile sensing and on device processing highlights both the feasibility of fast inference and the need for privacy-preserving pipelines~\cite{valliappan2020accelerating}. Our work adds evidence that lightweight models can be competitive for both gesture recognition and user identification using only ARKit transformation streams.

\section{Methodology}
We designed a data collection pipeline and a evaluation framework to assess the gaze gestures on devices. The system captures raw sensor data, preprocesses it into temporal sequences, and feeds it into models.

\subsection{Participants Recruitment and Ethics}
Participants were recruited via student and staff mailing lists at the first author's university using a recruitment advertisement. The pilot study included 4 participants (2 female, 2 male; mean age = 27.5, SD = 5.32). 2 participants reported having prior experience with eye-tracking or gaze-based interfaces, while the remaining were novices. Eligibility criteria included being at least 18 years old and having sufficient English proficiency to understand the study instructions. Furthermore, we obtained institutional ethical approval for collecting participant gaze-related signals. To reduce ethical risk, we log only transformation matrices and event metadata for analysis; no face images or videos are required for the reported experiments and thus deleted immediately after transformation. All participants provided informed consent, and data were stored and analysed under anonymized participant IDs.

\subsection{User Study Setup}

\subsubsection{Mobile Sensing System}
We built an iOS application using SwiftUI and ARKit to capture gaze performance. The app logs data at approximately 60\,Hz using the front-facing TrueDepth camera. We capture Head Pose and Eye Pose, both represented as 4$\times$4 transformation matrices describing the rotation and translation of the head and eyes relative to the camera in 3D space. An \textit{EventLog} simultaneously records gesture stages to ensure accurate data segmentation. In this controlled study, gesture onset and offset are explicitly defined by the protocol UI and recorded in the EventLog (rather than inferred from passive gaze behaviour), which isolates recognition performance from Midas-touch ambiguity during benchmarking. For real-world deployment, onset/offset detection should be paired with an explicit activation strategy (e.g., a mode switch, bounded activation region, or confirmatory trigger), which remains future work.

\subsubsection{Gesture Set}

\begin{figure}[t]
  \centering
  \includegraphics[width=0.95\linewidth]{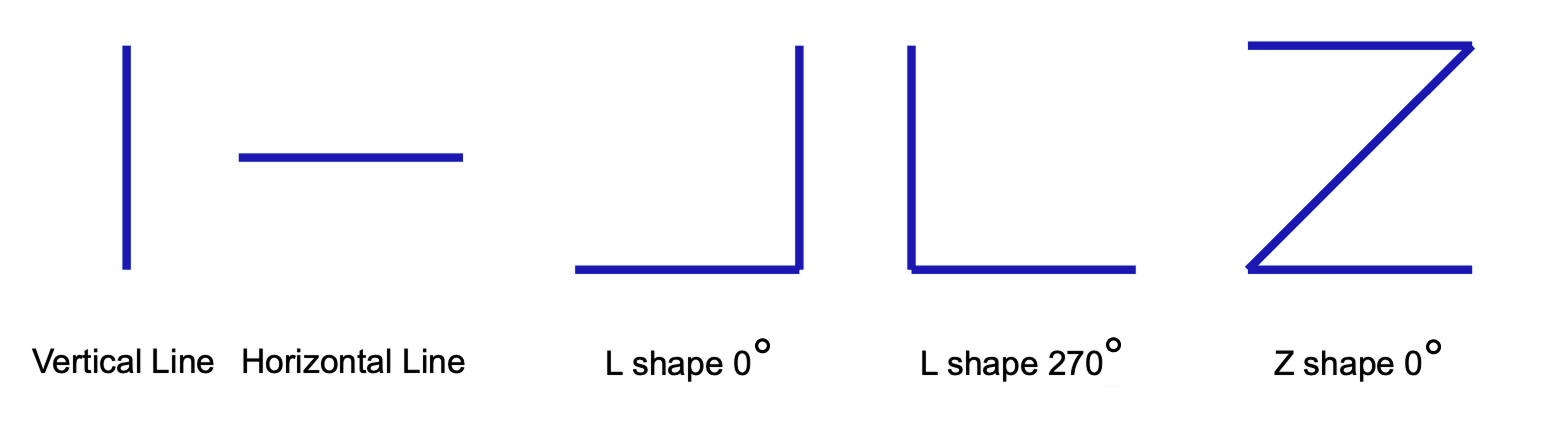}
  \caption{Gesture set. Five gaze-gesture patterns with increasing directional complexity: Vertical, Horizontal, L-shape ($0^\circ$), L-shape ($270^\circ$), and Z-shape ($0^\circ$). These templates are simple to learn yet sufficiently distinct to evaluate recognition under noisy mobile sensing.}
  % \Description{Illustrations of five gaze gestures: vertical line, horizontal line, two L-shapes with different orientations, and a Z-shape.}
  \Description{Illustration of the five gaze gestures used in the study, arranged as simple line and shape templates with increasing complexity: a vertical line, a horizontal line, an L-shape at 0 degrees, an L-shape at 270 degrees, and a Z-shape at 0 degrees. The set spans straight strokes, single-turn gestures, and a multi-segment gesture to test recognition under noisy mobile sensing.}
  \label{fig:gestures}
\end{figure}

We evaluate five gaze-gesture patterns from the gaze gesture dataset~\cite{lei2026people}, spanning a controlled range of spatial and directional complexity: \textit{Vertical Line, Horizontal Line, L-shape 0$^\circ$, L-shape 270$^\circ$, and Z-shape 0$^\circ$} (Figure~\ref{fig:gestures}). These specific directional strokes and shapes were selected based on established gesture taxonomies~\cite{lei2026people} and prior gaze-gesture research~\cite{drewes2007interacting}. The set was chosen to balance (i) \emph{learnability} and low visual complexity for novice users with (ii) \emph{directional distinctiveness} for robust classification under noisy mobile sensing. Additionally, it features (iii) \emph{increasing structural complexity} (straight strokes $\rightarrow$ single-turn shapes $\rightarrow$ multi-segment shapes) to probe whether performance degrades with added corners and direction changes. Finally, we selected shapes that fit comfortably within a phone display and can be repeatedly executed in a seated mobile setup without requiring large off-screen excursions.

\subsubsection{Study Protocol}
We conducted a controlled user study with 4 participants seated approximately 40\,cm from an iPhone~15~Pro~Max (running iOS~18). ARKit logged head and eye transformation matrices at a rate of $\sim$60\,Hz. The gesture configuration was standardized for all trials: patterns were set to the maximum size (300 points), centered on the screen, with a dot animation speed of 100 points per second. The study utilized a within-subjects design where each participant performed 5 distinct gestures across the four scaffolding stages (Follow, Fixed, IRecall, Recall). With 3 repetitions per gesture per stage, we collected a total of 60 trials per participant. To isolate the gesture recognition performance and prevent the Midas touch problem during data collection, trial initiation and termination were explicitly demarcated by the user via a screen tap, serving as a reliable segmentation trigger. Future real-world deployments could adapt this using a bezel swipe or dwell-based trigger.

To ensure users can reliably learn and reproduce gaze gestures, we have designed a learning-theory-grounded protocol based on \textit{scaffolding} and \textit{retrieval practice}~\cite{van2010scaffolding,roediger2011critical}. Scaffolding reduces cognitive load during initial encoding, while retrieval practice strengthens motor learning through active recall.

\begin{figure}[t]
  \centering
  \includegraphics[width=0.8\linewidth]{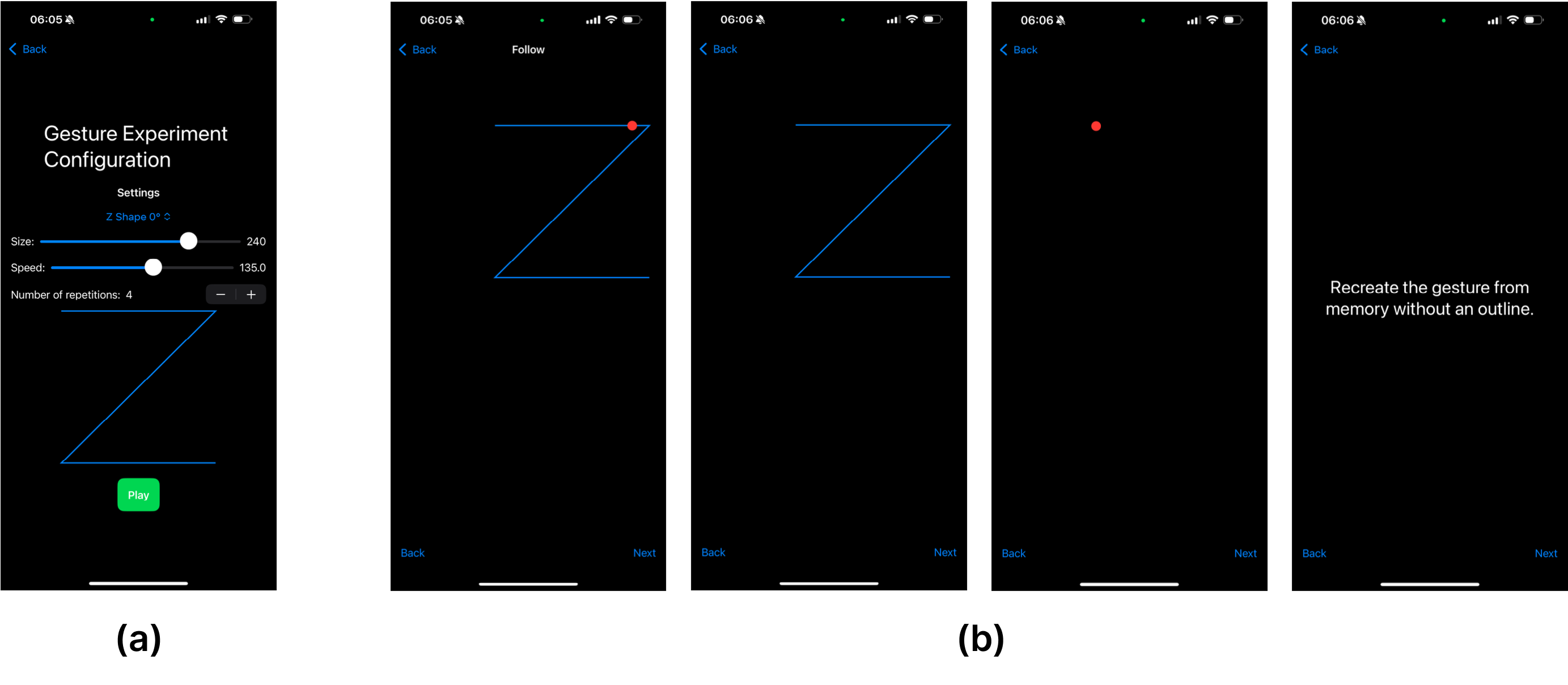}
  \caption{Scaffolded study protocol. (a) Gesture configuration UI for adjusting parameters (e.g., size, speed, repetitions). (b) Four-stage guidance-to-recall sequence: Follow (moving guide), Fixed (static outline), IRecall (start-point cue), and Recall (no visual guidance), designed to transition from low-variance tracing to memory-based production.}
  % \Description{Screenshots of the TinyGaze app: a configuration screen and four study stages that gradually remove visual guidance from a moving guide to a blank screen.}
  \Description{Composite figure showing the TinyGaze study interface and scaffolded protocol. Panel (a) shows the gesture configuration screen, where parameters such as pattern size, speed, and repetitions can be adjusted. Panel (b) shows four study stages that progressively reduce visual guidance: Follow with a moving guide dot, Fixed with a static gesture outline, IRecall with only a start-point cue, and Recall with no visual guidance. The sequence illustrates a transition from guided tracing to memory-based gesture production.}
  \label{fig:study-protocol}
\end{figure}

Participants are requested to perform each gesture repeatedly across four stages with progressively reduced visual guidance (Figure~\ref{fig:study-protocol}). In \textit{Follow} (fully guided), users trace a moving dot to acquire the spatiotemporal rhythm and trajectory. In \textit{Fixed} (static scaffold), users trace a stationary outline, shifting from pursuit to self-paced execution while retaining a strong spatial cue. As guidance fades, \textit{IRecall} (faded cue) shows only the starting point, requiring internal reconstruction with a minimal anchor. In \textit{Recall} (free retrieval), users perform the gesture from memory on a blank screen, approximating real-world use without visual guides. This design creates a continuum of variability: early stages yield low-variance, guidance-driven data, while Recall yields higher-variance, memory-driven data. Evaluating across stages tests whether models capture \emph{gesture identity} rather than artifacts of the guidance display.

\subsection{Preprocessing and Evaluation Protocol}
\noindent\textit{Preprocessing.} Raw ARKit logs are processed via a pipeline that extracts gesture intervals from the EventLog and synchronizes head/eye streams. To ensure consistency, we resample each performance to a fixed length of $T{=}64$ frames. For features, we flatten the 4$\times$4 transformation matrices\footnote{\url{https://developer.apple.com/documentation/arkit/arfaceanchor}} indicating the position and orientation of eyes and head, concatenating them to form three input modalities: \textit{Head-only} (16D), \textit{Eye-only} (32D), and \textit{Eye+Head} (48D).

\noindent\textit{Models.} We evaluate \textit{TinyHAR}~\cite{zhou2022tinyhar}, a lightweight compact model, against two deep baselines: \textit{DeepConvLSTM}~\cite{wan2019multivariate} and \textit{SA-HAR}~\cite{mahmud2020human}. All models are trained using a 1.5\,s sliding window with 50\% overlap (training) and 90\% overlap (testing).

\noindent\textit{Validation Protocol.} We assess two tasks with distinct cross-validation strategies to prevent leakage: (i) \textit{Gesture Recognition (5 classes):} Evaluated via \textit{Leave-One-Subject-Out (LOSO)} to test generalizability to unseen users; (ii) \textit{User Identification (4 classes):} Evaluated via \textit{Stratified 4-Fold CV}, splitting by \textit{trials} (not windows) to ensure robust biometric verification. 
We also report Macro F1 scores to account for any class imbalances.

\section{Results}
We evaluated the models based on classification performance and computational efficiency. We further analysed the impact of sensor modality to understand the drivers of recognition accuracy.

\textbf{Learnability Outcomes.} 
While our primary evaluation focuses on model accuracy, the scaffolded protocol also yielded measurable improvements in gesture recall. Participants demonstrated a 98\% success rate during the guided \textit{Follow} and \textit{Fixed} stages. As visual cues faded, initial error rates in the \textit{IRecall} stage were 15\%, primarily due to spatial scaling issues, but stabilized to a 92\% successful completion rate in the final free \textit{Recall} stage. This confirms that the distinctiveness of the gestures holds even when visual guides are completely removed.

\textbf{Gesture Recognition.}
Table~\ref{tab:gesture-rec} reports performance on the combined multi-participant dataset (P0123-G). TinyHAR achieves the best Macro F1 (0.960), outperforming both DeepConvLSTM (0.762) and SA-HAR (0.853).  TinyHar has  far fewer parameters (22.8$\times$ smaller than DeepConvLSTM; 9.5$\times$ smaller than SA-HAR). This suggests that compact inductive biases (and fewer parameters) can generalize better than deeper models in noisy ARKit-based sensing.

\begin{table}[t]
  \centering
  \caption{Gesture recognition performance on P0123-I.}
  \Description{A four-column table comparing gesture recognition performance across three models on dataset P0123-I. Columns are Model, Macro F1, Params, and Execution Time in milliseconds. TinyHAR achieves the highest Macro F1 of 0.960 with 46,094 parameters and 40.0 ms execution time. DeepConvLSTM achieves 0.762 Macro F1 with 1,050,208 parameters and 39.4 ms. SA-HAR achieves 0.853 Macro F1 with 440,412 parameters and 77.3 ms. The table shows that TinyHAR has the best recognition performance with far fewer parameters than the deeper baselines.}
  \label{tab:gesture-rec}
  \setlength{\tabcolsep}{3pt}
  \begin{tabular}{@{}lccc@{}}
    \toprule
    \textbf{Model} & \textbf{Macro F1} & \textbf{Params} & \textbf{Exec.\ Time (ms)} \\
    \midrule
    TinyHAR~\cite{zhou2022tinyhar} & 0.960 & 46,094 & 40.0 \\
    DeepConvLSTM~\cite{wan2019multivariate} & 0.762 & 1,050,208 & 39.4 \\
    SA-HAR~\cite{mahmud2020human} & 0.853 & 440,412 & 77.3 \\
    \bottomrule
  \end{tabular}
\end{table}

\textbf{User Identification.}
Table~\ref{tab:user-id} summarizes overall user identification performance on P0123-I. TinyHAR achieves near-ceiling performance in this pilot setting. However, given the small sample size ($N{=}4$) and single-session controlled collection, these results should be interpreted cautiously. To assess robustness across gesture types, Table~\ref{tab:tinyhar-userid-by-gesture} reports TinyHAR results when training/testing on each single-gesture subset, alongside the full authentication set. Performance is near-ceiling for most gestures (Macro F1 $=1.000$), with a modest drop for the L$270^\circ$ subset (Macro F1 $=0.937$), indicating that some trajectories may carry weaker identity cues under this sensing setup. Finally, the modality ablation for \emph{gesture recognition} (Table~\ref{tab:tinyhar-modality-compact}) shows that head pose alone is often highly informative, while fusing eye and head signals yields the most robust performance on average, suggesting complementary contributions from eye features in the full pipeline.

\begin{table}[t]
  \centering
  \caption{User identification on P0123-I. TinyHAR achieves near-optimal performance while using far fewer parameters and reduced compute relative to deep baselines.}
  \Description{A three-column table summarizing user identification performance on dataset P0123-I for three models. Columns are Model, Macro F1, and Execution Time in milliseconds. TinyHAR achieves a Macro F1 of 0.997 with 190.1 ms execution time. DeepConvLSTM achieves 1.000 Macro F1 with 329.2 ms, and SA-HAR achieves 0.961 Macro F1 with 513.0 ms. The table indicates near-ceiling performance for TinyHAR in this pilot setting while using less compute time than the deeper baselines.}
  \label{tab:user-id}
  \begin{tabular}{lcc}
    \toprule
    \textbf{Model} & \textbf{Macro F1} & \textbf{Exec.\ Time (ms)} \\
    \midrule
    TinyHAR~\cite{zhou2022tinyhar} & 0.997 & 190.1 \\
    DeepConvLSTM~\cite{wan2019multivariate} & 1.000 & 329.2 \\
    SA-HAR~\cite{mahmud2020human} & 0.961 & 513.0 \\
    \bottomrule
  \end{tabular}
\end{table}

\textbf{Additional observations (Modality and Variability).}
Beyond the main results, we analysed the performance variation across different sensor inputs and on different subjects. As detailed in Table~\ref{tab:tinyhar-modality-compact}, the Head-only and Eye\_head yielded the highest performance overall (0.816 and 0.824). This suggests that users heavily rely on head movement to trace large on-screen patterns on screen, making head pose important. In addition, performance varied across individuals; for example, P3 achieved macro F1 as high as $\sim$0.928 while P1 produced as low as $\sim$0.301, highlighting the need for personalized calibration.

\begin{table}[t]
  \centering
  \caption{User identification across gesture subsets and full set.}
  \Description{A four-column table reporting TinyHAR user identification performance across gesture subsets and the full authentication set. Columns are Gesture type, Accuracy, Weighted F1, and Macro F1. The full set labelled Auth (all) has 0.997 for all three metrics. Vertical, Horizontal, L 0, and Z 0 each achieve perfect scores of 1.000. The L 270 subset shows lower performance, with Accuracy 0.938, Weighted F1 0.937, and Macro F1 0.937. The table shows that user identification is good for most gesture types, with L 270 being the most challenging subset.}
  \label{tab:tinyhar-userid-by-gesture}
  \begin{tabular}{lccc}
    \toprule
    \textbf{Gesture type} & \textbf{Accuracy} & \textbf{Weighted F1} & \textbf{Macro F1} \\
    \midrule
    Auth (all) & 0.997 & 0.997 & 0.997 \\
    Vertical   & 1.000 & 1.000 & 1.000 \\
    Horizontal & 1.000 & 1.000 & 1.000 \\
    L 0        & 1.000 & 1.000 & 1.000 \\
    L 270      & 0.938 & 0.937 & 0.937 \\
    Z 0        & 1.000 & 1.000 & 1.000 \\
    \bottomrule
  \end{tabular}
\end{table}

\begin{table}[t]
  \centering
  \caption{Gesture recognition modality ablation using TinyHAR (Macro F1).}
  \Description{A six-column table showing TinyHAR gesture recognition Macro F1 scores for different sensor input modalities across participants P0, P1, P2, and P3, plus an average column. Rows correspond to Eye_Head, Eyes, Left_Eye, Right_Eye, and Head inputs, followed by an overall average row. Eye_Head has the highest average performance at 0.824, followed by Head at 0.816. Eyes, Left_Eye, and Right_Eye achieve lower averages of 0.655, 0.633, and 0.649, respectively. Participant-level scores vary substantially, with P1 lower than the others across most modalities. The table indicates that head pose is highly informative and that fusing head and eye signals yields the strongest average performance.}
  \label{tab:tinyhar-modality-compact}
  \begin{tabular}{lccccc}
    \toprule
    \textbf{Input} & \textbf{P0} & \textbf{P1} & \textbf{P2} & \textbf{P3} & \textbf{Avg}\\
    \midrule
    \texttt{Eye\_Head}  & 0.796 & 0.683 & 0.928 & 0.889 & 0.824\\
    \texttt{Eyes}       & 0.834 & 0.469 & 0.659 & 0.659 & 0.655\\
    \texttt{Left\_Eye}  & 0.699 & 0.301 & 0.699 & 0.831 & 0.633\\
    \texttt{Right\_Eye} & 0.671 & 0.308 & 0.784 & 0.834 & 0.649\\
    \texttt{Head}       & 0.960 & 0.506 & 0.879 & 0.919 & 0.816\\
    \midrule
    \textbf{Avg}        & 0.792 & 0.453 & 0.790 & 0.826 & 0.715 \\
    \bottomrule
  \end{tabular}
\end{table}

\section{Discussion}
The staged protocol (Follow$\rightarrow$Recall) demonstrated the importance of evaluating learnability, ensuring the system recognizes internalised motor patterns rather than merely guided tracing. While the near-perfect user identification scores (Macro F1 = 0.997) highlight the potential of gaze as a behavioural biometric, we explicitly scope this as a pilot feasibility study. The small sample size ($N = 4$) is a significant limitation. The high accuracy may partly reflect overfitting to session-specific confounds, such as stable biometric cues, rather than robust, persistent gesture dynamics. Furthermore, the high predictive power of the head-only dataset implies that mobile ``gaze gestures'' are effectively embodied head-eye coordination~\cite{hu2025hoigaze}. Users naturally stabilize their eyes relative to the device or adopt coupled motion to reduce ocular effort. While treating head motion as a primary signal improves recognition, it introduces practical constraints. Relying heavily on head movement can lead to physical fatigue over prolonged use and may suffer from low social acceptability in public spaces. Consequently, interaction designs must balance this head-eye coupling, and future evaluations must explicitly test performance in scenarios where head movement is physically or socially constrained.

Our preliminary results demonstrate a promising direction for real-time on-device gesture recognition. In this controlled pilot benchmark, the compact TinyHAR model achieves the strongest gesture-recognition performance (Macro F1 = 0.960) with substantially fewer parameters than the deeper baselines. Specifically, TinyHAR required only 8.5 ms for inference per sliding window on the gesture task (excluding offline training costs), compared to 18.2 ms for SA-HAR, effectively halving the latency overhead. In the more complex authentication task, this inference efficiency gap widened (22.4 ms vs 65.8 ms). This combination of high accuracy and sub-100ms inference latency supports the feasibility of always-on gaze sensing directly on mobile processors, offering a favorable trade-off between sampling rate and battery energy impact. To validate generalizability, future work can scale beyond this pilot’s controlled, single-session environment. Key priorities include collecting multi-session data to test temporal stability, expanding to ``in-the-wild'' conditions (e.g., walking, varying illumination), and investigating personalization techniques to refine both gesture recognition and authentication on resource-constrained devices.

Based on the results, we summarize the following three key suggestions for gesture implementation:

\noindent\textbf{1. Gesture design for learnability (scaffold $\rightarrow$ recall):} Use staged guidance (Follow $\rightarrow$ Fixed $\rightarrow$ IRecall $\rightarrow$ Recall) to help users internalize gesture trajectories.

\noindent\textbf{2. Treat head motion as a first-class signal:} Mobile ``gaze gestures'' often involve embodied head-eye coordination. Head pose is a core signal and should not be simply normalized through gaze estimation pipelines.

\noindent\textbf{3. Prefer compact models for deployment:} Lightweight time-series architectures can match strong accuracy with a smaller parameter budget, making them better suited to on-device inference under mobile contexts.

\section{Conclusion}
We presented an end-to-end pipeline for collecting and recognizing mobile gaze gestures using commodity ARKit head/eye transforms, together with a scaffolded guidance-to-recall protocol designed to elicit recall-ready gestures. In a controlled pilot evaluation, a compact HAR-style model (TinyHAR) achieved strong performance for both gesture recognition (Macro F1 $=0.960$) and user identification (Macro F1 $=0.997$) while using far fewer parameters than deeper baselines. Our modality analysis suggests head pose is a strong contributor to recognition accuracy, motivating future designs that explicitly account for embodied head--eye coordination. Future work should (i) report deployment-grade latency and energy profiling, (ii) evaluate multi-session temporal stability, and (iii) validate robustness in in-the-wild mobile conditions.

\begin{acks}
We thank all participants for their time and contributions. We also thank the CHI reviewers for their constructive feedback, which helped improve the clarity and presentation of this work. Finally, we are grateful to our colleagues for helpful discussions and support throughout the project.
\end{acks}

% \begin{acks}
% \end{acks}

\bibliographystyle{ACM-Reference-Format}
\bibliography{reference}


\clearpage
\appendix

\end{document}